# Superconducting four-fold Fe(Te,Se) film on six-fold magnetic MnTe via hybrid symmetry epitaxy

*Xiong Yao, *,† Alessandro R. Mazza,⊥ Myung-Geun Han,∥ Hee Taek Yi, § Deepti Jain,§ Matthew Brahlek,⊥ and Seongshik Oh*,†*

†Center for Quantum Materials Synthesis and Department of Physics & Astronomy, Rutgers, The State University of New Jersey, Piscataway, New Jersey 08854, United States

⊥ Materials Science and Technology Division, Oak Ridge National Laboratory, Oak Ridge, Tennessee 37831, United States

∥ Condensed Matter Physics and Materials Science, Brookhaven National Laboratory, Upton, New York 11973, United States

§Department of Physics & Astronomy, Rutgers, The State University of New Jersey, Piscataway, New Jersey 08854, United States

*Email: xiong.yao@rutgers.edu

*Email: ohsean@physics.rutgers.edu

Phone: +1 (848) 445-8754 (S.O.)

ABSTRACT

Epitaxial Fe(Te,Se) thin films have been grown on various substrates but never been realized on magnetic layers. Here we report the epitaxial growth of four-fold Fe(Te,Se) film on a six-fold antiferromagnetic insulator, MnTe. The Fe(Te,Se)/MnTe heterostructure shows a clear superconducting transition at around 11 K and the critical magnetic field measurement suggests the origin of the superconductivity to be bulk-like. Structural characterizations suggest that the uniaxial lattice match between Fe(Te,Se) and MnTe allows a hybrid symmetry epitaxy mode, which was recently discovered between Fe(Te,Se) and $Bi_2Te_3$. Furthermore, Te/Fe flux ratio during deposition of the Fe(Te,Se) layer is found to be critical for its superconductivity. Now that superconducting Fe(Te,Se) can be grown on two related hexagonal platforms, $Bi_2Te_3$ and MnTe, this result opens a new possibility of combining topological superconductivity of Fe(Te,Se) with the rich physics in the intrinsic magnetic topological materials $(MnTe)_n(Bi_2Te_3)_m$ family.

Keywords: Hybrid symmetry epitaxy, Superconductor/magnetic insulator heterostructure, Fe(Te,Se)/MnTe, Uniaxial lattice match, Superconductivity

Topological superconductors (TSCs), as one of the candidate platforms for realizing Majorana fermions and fault-tolerant quantum computing, have long been a pursuit for condensed matter physicists.[1] Chemically intercalated topological insulators (TIs) and superconducting proximity heterostructures are two general approaches to implementing TSCs.[2-10] Recently, superconducting states with Majorana zero modes have been observed in an iron-based superconductor Fe(Te,Se) (FTS) with $T_C$ up to 14.5 K, using angle resolved photoemission spectrocopy[11] and scanning tunneling spectroscopy.[12-14]

FTS thin films have been intensively investigated since the discovery of superconductivity in bulk FTS crystals. In general, epitaxial growth of FTS films requires the substrates to have the same four-fold in-plane symmetry as that of FTS, so MgO, SrTiO$_3$, LaAlO$_3$, LSAT and CaF$_2$ substrates have been commonly used to grow superconducting FTS films .[15-21] Recently, however, it was discovered that superconducting FTS films can be epitaxially grown on a three-fold topological insulator material, Bi$_2$Te$_3$, due to the rare uniaxial lattice match between FTS and Bi$_2$Te$_3$, a growth mode dubbed "hybrid symmetry epitaxy".[22, 23] Here, we show that similar hybrid symmetry epitaxy mode allows FTS to grow epitaxially on six-fold MnTe layer. It is notable that MnTe is an antiferromagnetic (AFM) insulator with a Neel temperature of ~310 K and plays an important role in the recently discovered intrinsic magnetic topological insulator family (MnTe)$_m$(Bi$_2$Te$_3$)$_n$.[24-26] This is the first time that superconducting FTS film is grown on a magnetic layer.

We grew the FTS/MnTe films on 10 mm × 10 mm Al$_2$O$_3$ (0001) substrates using a custom built SVTA MOS-V-2 molecular beam epitaxy (MBE) system with a base pressure of low $10^{-10}$ Torr. To improve the lattice matching between Al$_2$O$_3$ and MnTe, we grew 5 nm-thick In$_2$Se$_3$ as a buffer layer, which involves a multistep growth strategy that uses Bi$_2$Se$_3$ as a seed layer: details can be found in our previous reports.[27, 28] After In$_2$Se$_3$ buffer layer was grown at 300 °C and raised to 600 °C to remove the Bi$_2$Se$_3$ seed layer, the substrate was cooled down to 450 °C for the growth of MnTe layer. Then, the FTS layer was grown at 350°C on MnTe by opening the shutters of Fe and Te with preadjusted fluxes. Because our chamber is used to grow both Se and Te compounds and that Se tends to stick more easily to Fe than Te does, all the FeTe films we have used in this study have Se impurities incorporated, at the level of a few percent according to Rutherford backscattering spectroscopy (RBS) analysis. So even though all the films used in this study are nominally FeTe, they are still called as FTS.

Figure 1a-b show the reflection high-energy electron diffraction (RHEED) patterns of the MnTe layer and the FTS overlayer grown on top. The bright streaky features in Figure 1a-b imply the high-quality epitaxial growth of the MnTe and FTS layers. The in-plane lattice symmetries of these two layers can be confirmed by the spacing ratio of the RHEED streaks in two high-symmetry directions. For MnTe layer, the lattice symmetry is six-fold, which is consistent with the spacing ratio of $\sqrt{3}$ marked in Figure 1a. Similarly, four-fold FTS layer shows a spacing ratio of $\sqrt{2}$ in Figure 1b. As indicated by the red dash lines connecting Figure 1a and b, the RHEED spacings of MnTe and FTS layers are the same in one high symmetry direction, despite their different in-plane lattice symmetries. The high-quality epitaxial growth of FTS overlayer is very similar to the novel "hybrid symmetry epitaxy" mode, which was recently reported in FTS/Bi$_2$Te$_3$ heterostructures by our group.[22] The in-plane lattice spacing of the MnTe layer is 4.38 Å based on the RHEED spacing. As shown in Figure 1c, the lattice of MnTe and FTS (lattice constant 3.79 Å) can match along one axis due to the $2:\sqrt{3}$ relative ratio between their lattice spacings, i.e., the uniaxial lattice matching. The uniaxial lattice matching was shown to be a critical factor for hybrid symmetry epitaxy of the four-fold FTS films on a three or six-fold platform in our previous

report.[22] The characteristic RHEED patterns of the FTS layer appear right after opening of the shutters, suggesting a smooth growth transition at the interface, which is also verified by the high-angle annular dark-field scanning tunneling electron microscopy (HAADF-STEM) image in Figure 1d. The well-defined, sharp interface between FTS and MnTe shown in Figure 1d confirms the high-quality of the heterostructure, enabled by the uniaxial lattice match.

Figure 2a and b give the X-ray diffraction (XRD) 2θ scan and azimuthal φ scans of an FTS/MnTe sample, respectively. All the MnTe and FeTe peaks in Figure 2a can be identified with c-axis oriented (00n) peaks of each phase. Like the FTS/$Bi_2Te_3$ heterostructure,[22] the FTS top layer exhibits twelve-fold peaks in the in-plane φ scans even though the MnTe bottom layer shows six-fold peaks. The twelve-fold peaks of FTS is consistent with the twelve-fold rotational symmetry of RHEED patterns of the FTS layer, which can be explained by in-plane twins of the four-fold FTS layer on top of the three crystallographically equivalent orientations of the MnTe base layer, details can be found in the previous report.[22] Notably, the peaks of MnTe layer and FTS layer are well aligned, as shown in the right panel of Figure 2b. This implies that the stacking alignment between MnTe and FTS is completely consistent with the schematic layout in Figure 1c.

The FTS/MnTe heterostructure is an integration of an iron-based superconductor and an AFM insulator. It would be insightful to investigate the superconducting properties of this heterostructure. Figure 3a gives the temperature-dependent longitudinal resistance of an FTS/MnTe sample from 300 K to 2 K. The upturn behavior around 55 K is believed to be the signature of AFM transition in the FTS layer rather than that of the MnTe layer considering that the Neel temperature of the MnTe system is known to be higher than 300 K.[29, 30] It is also notable that such a hump feature is absent in the FTS/$Bi_2Te_3$ system.[22] This difference can be partly explained by the fact that MnTe is insulating whereas $Bi_2Te_3$ is conducting and, thus, can mask such a hump-like feature even if it were there. However, one cannot completely rule out the possibility that the AFM transition of the FTS layer is absent on $Bi_2Te_3$ for some unknown reason, and this is an interesting question for future studies. As shown in the inset of Figure 3a, the critical temperature determined by the intersection point of linear extrapolations from normal state and transition region is 11 K, slightly lower than the values reported in FTS/$Bi_2Te_3$.[22] The transition is broader than FTS/$Bi_2Te_3$ heterostructure and residual resistance is observed over a range of temperature before reaching zero resistance. Considering that the anion of FTS/$Bi_2Te_3$ and FTS/MnTe is the same, the lower $T_C$ and broader transition in FTS/MnTe is likely related to Mn. This broad transition could be either due to magnetic/superconducting proximity coupling between FTS and MnTe layers or due to inter-diffusion between Fe and Mn atoms during the growth, both of which can weaken the superconductivity of FTS.

To investigate the anisotropy of superconductivity in FTS/MnTe, the temperature-dependent longitudinal resistance of the same sample was measured under magnetic fields perpendicular and parallel to the ab-plane, respectively, as shown in Figure 3b and c. Figure 3d shows the upper critical fields along the two directions determined by 50% of normal state resistance in Figure 3b and c. Both $H_{C2}^{\perp}(T)$ and $H_{C2}^{//}(T)$ show sublinear temperature dependence rather than the linear $H_{C2}^{\perp}(T)$ dependence that has been commonly observed in 2D superconductors, indicating that the observed superconductivity has significant bulk contribution. Although superconductivity in $Bi_2Te_3$/FeTe and $Sb_2Te_3$/$Fe_{1+y}$Te has been reported to be dominated by the interfacial origin,[29-31] superconductivity in FTS/$Bi_2Te_3$ seems to be dominated by bulk, rather than interface, superconductivity.[22] As in the case of FTS/$Bi_2Te_3$, even though we cannot completely rule out the presence of interfacial superconductivity in FTS/MnTe, the critical field measurements are more consistent with bulk dominant superconductivity in this system. Further investigations will be needed to fully reveal the origin of the superconductivity in this system.

Another notable observation is that the Te/Fe flux ratio is a critical factor determining the superconducting nature of this system. In order to study this, we fixed the thickness of the FTS/MnTe heterostructures and only changed the Te/Fe ratio during the deposition of FTS layer. Figure 4 gives the superconducting properties of samples with varying Te/Fe ratios. Superconductivity was observed in Figure 4a and b for samples with Te/Fe flux ratios of 3.6, 4.0, 5.1 and 8.0. The optimal Te/Fe ratio is around 4.0 based on the critical temperatures and transition width of these samples. The degraded superconductivity in samples with Te/Fe ratio higher than 4.0 are likely related to the formation of $FeTe_2$ impurity phase. The $FeTe_2$ phase is reported to be a semiconductor and can easily form during a Te rich growth condition.[32, 33] With increasing Te/Fe flux ratio, the percentage of $FeTe_2$ phase will grow and gradually segregate the superconducting FTS islands, leading to the decrease of superconducting critical temperature with broadening transition. One the other hand, Te/Fe flux ratio lower than 4 is not favorable to superconductivity either, as shown in Figure 4b and c. Superconducting transition broadens at Te/Fe = 3.6 and completely vanishes for Te/Fe = 2.2 and 2.7. With decreasing Te/Fe ratio, the Te poor growth environment could lead to increased excess Fe, which can then collapse superconductivity. The AFM-transition-induced upturn becomes more prominent in samples with Te/Fe = 2.2 and 2.7 compared with other samples, exhibiting another signature of increased excess Fe.[30, 34] These excess Fe atoms that reside at interstitial sites can magnetically couple with adjacent Fe square-planar sheets and lead to charge carrier localization.[34] Excess Fe has also been reported to be deleterious for the superconductivity in $Sb_2Te_3/Fe_{1+y}Te$ heterostructure [30] as well as FTS bulk crystals.[35]

We have successfully grown four-fold FTS with superconducting $T_C$ of 11 K on six-fold MnTe. Structural characterizations show that uniaxial lattice match between FTS and MnTe is a key to the hybrid symmetry epitaxy. Critical field measurements suggest that the observed superconductivity is dominated by a bulk (rather than interface) contribution. Furthermore, Te/Fe flux ratio, which affects the formation of $FeTe_2$ phase as well as the level of excess Fe, is found to be a critical factor that determines the superconducting properties of the FTS/MnTe system. These results open a rich playground for investigating the interplay between topological superconductivity and magnetism in heterostructures composed of FTS and magnetic $(MnTe)_m(Bi_2Te_3)_n$ layers.

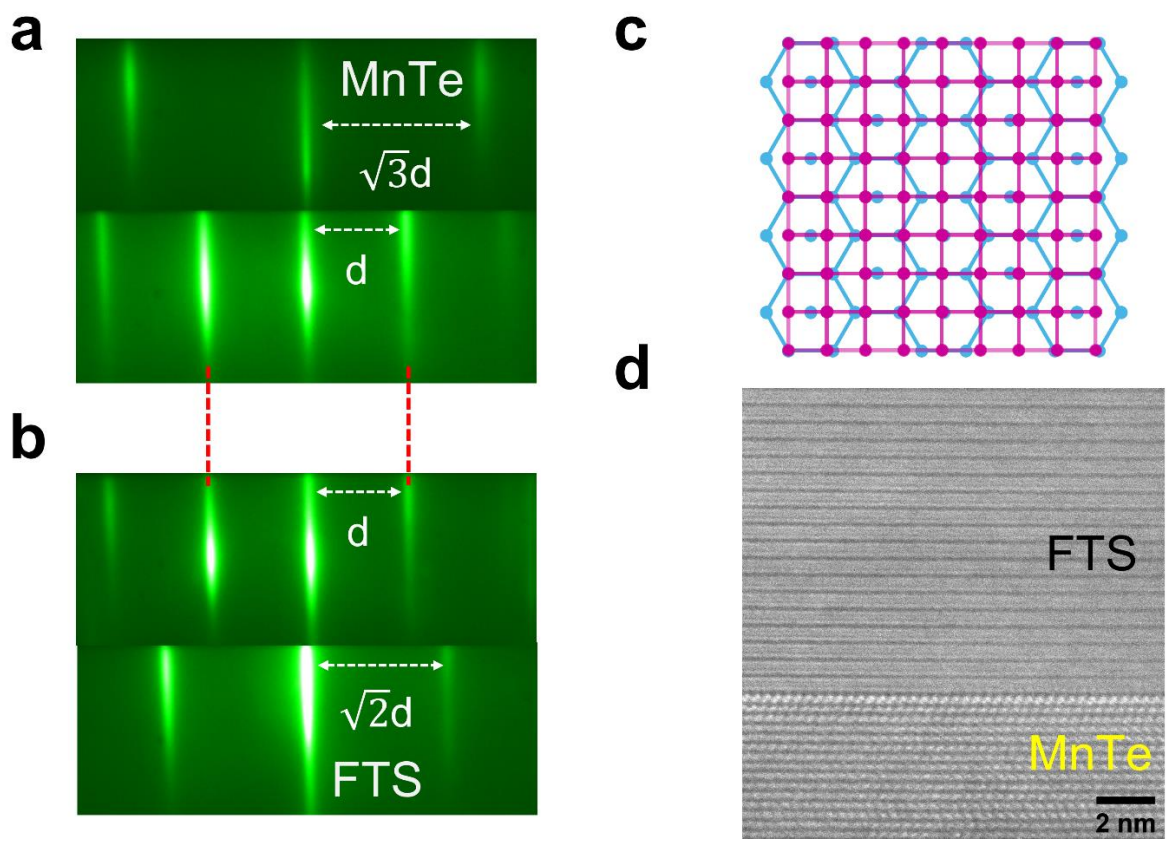

Figure 1. (a and b) Reflection high-energy electron diffraction (RHEED) patterns of (a) 10 nm MnTe and (b) 22 nm FTS grown on MnTe. The arrow marks indicate the RHEED streak spacings. The red dash guidelines indicate that the RHEED spacings are the same. (c) Schematic of FTS (purple) lattice overlaid on top of MnTe (blue) lattice: the dots represent Te(Se) atoms. (d) High-angle annular dark-field scanning transmission electron microscopy (HAADF-STEM) image for an FTS/MnTe sample.

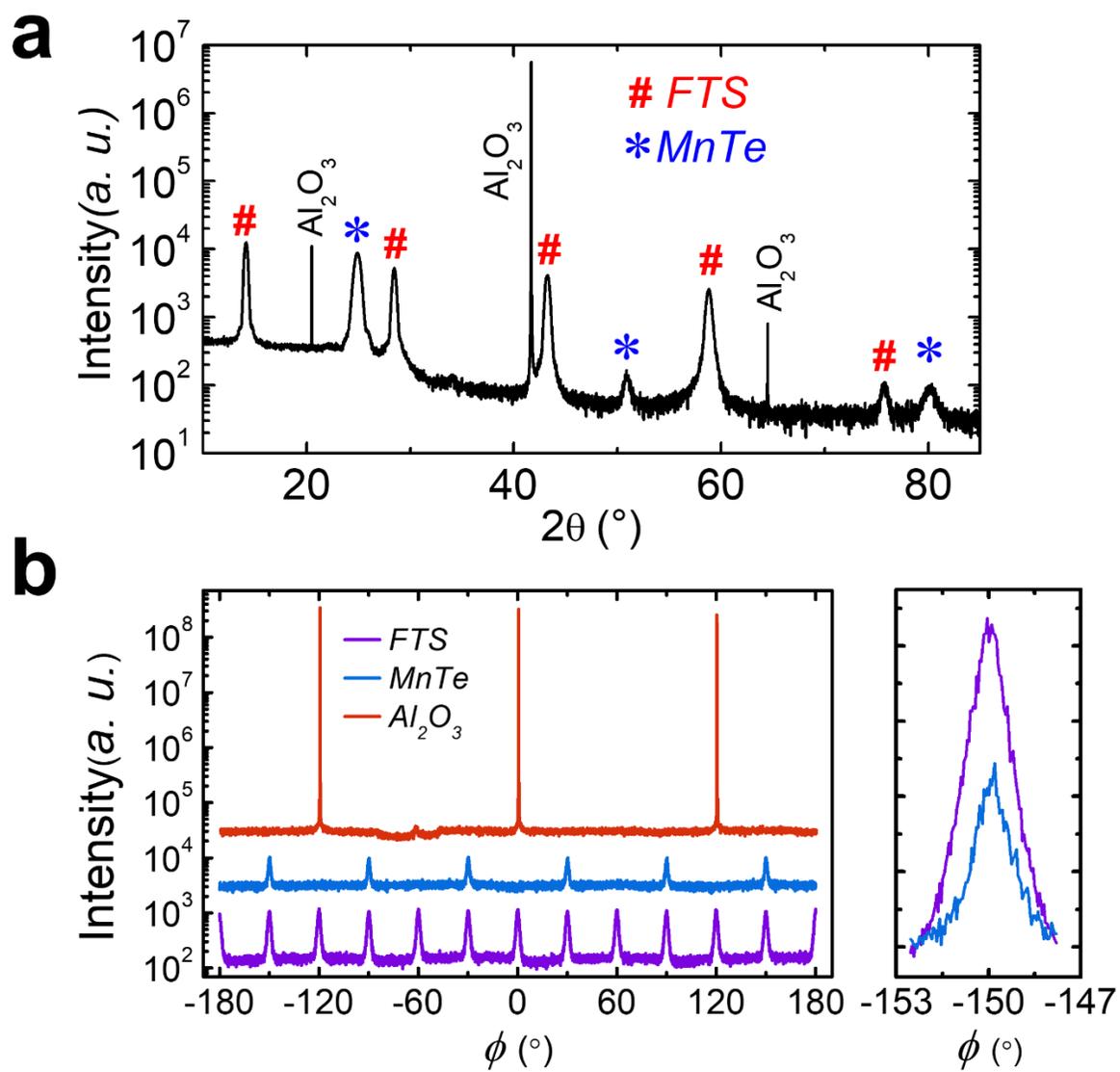

Figure 2. (a) XRD 2θ scan of an FTS (22 nm)/MnTe (10 nm) film. (b) In-plane XRD φ scans of the same sample in (a).

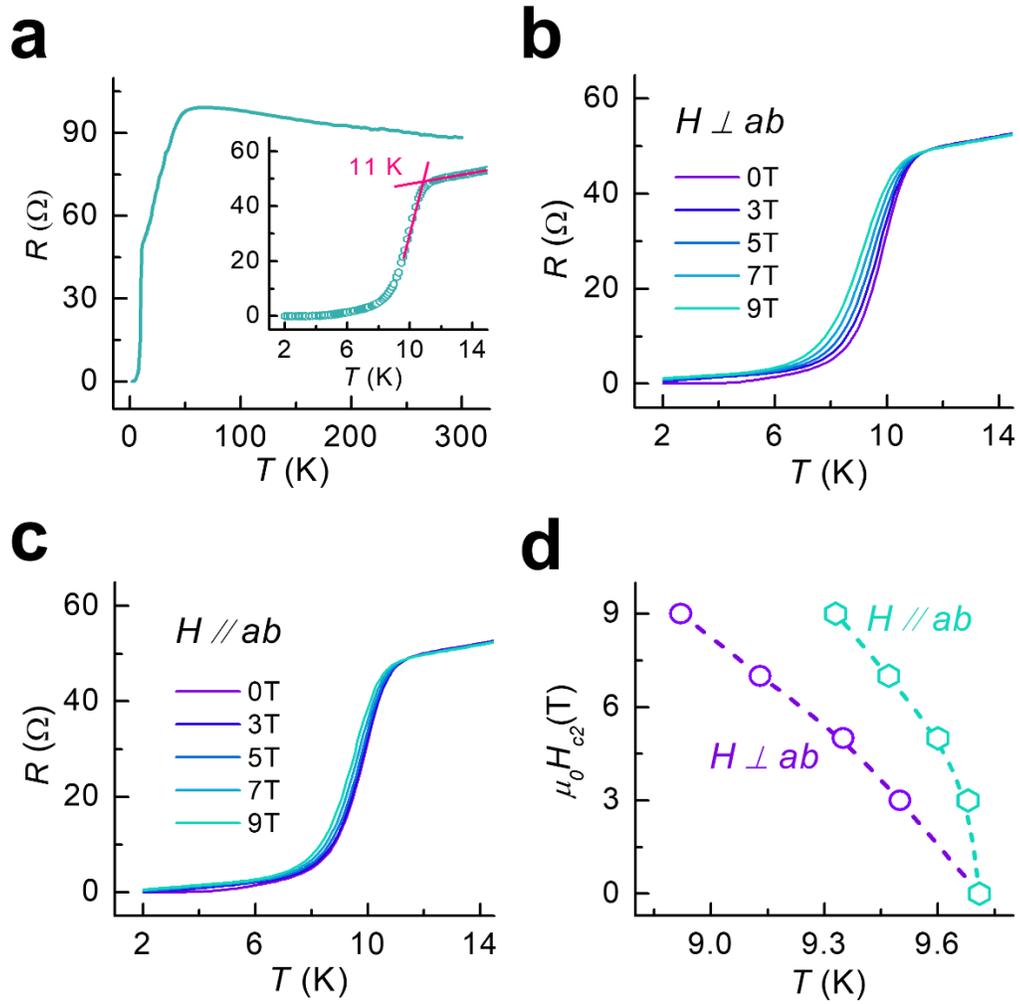

Figure 3. (a) Temperature-dependent longitudinal resistance of an FTS (22 nm)/MnTe (10 nm) film from 300 K to 2 K. Inset shows the enlarged plot from 2 K to 18 K. Intersection of linear extrapolations from normal-state and superconducting transition regions gives $T_C^{onset}$ of 11 K. (b and c) Temperature-dependent longitudinal resistance of the same sample in (a) under varying magnetic fields (b) perpendicular and (c) parallel to the ab plane. (d) $H_{C2}^{\perp}(T)$ and $H_{C2}^{//}(T)$ obtained by the data in Figure (b) and (c), determined by 50% $R_n$.

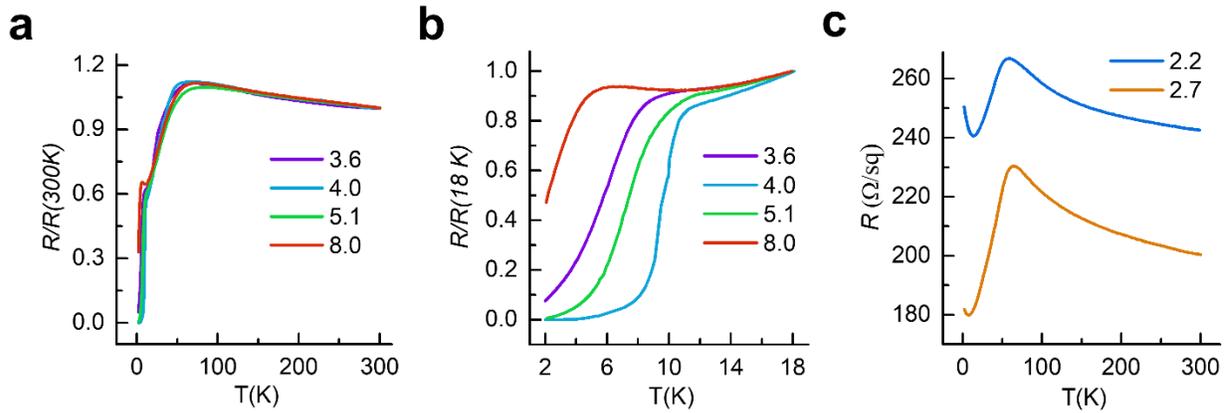

Figure 4. (a) Normalized longitudinal resistance of FTS (22 nm)/MnTe (10 nm) films with varying Te/Fe flux ratios. (b) The enlarged plot from 2 K to 18 K in (a). (c) Longitudinal sheet resistance of non-superconducting FTS/MnTe films with varying Te/Fe flux ratios.

AUTHOR INFORMATION

Corresponding Authors

*E-mail: xiong.yao@rutgers.edu

*E-mail: ohsean@physics.rutgers.edu

Author Contributions

X.Y. and S.O. conceived the experiments. X.Y., H.Y. and D.J. grew the thin films. X.Y. performed the transport measurements and analyzed the data with S.O. A.M. and M.B. performed the XRD measurements. M.H. performed the STEM measurements. X.Y. and S.O. wrote the manuscript with contributions from all authors.

Notes

The authors declare no competing financial interest.

ACKNOWLEDGMENT

The work at Rutgers is supported by National Science Foundation's DMR2004125, Army Research Office's W911NF2010108, MURI W911NF2020166, and the center for Quantum Materials Synthesis (cQMS), funded by the Gordon and Betty Moore Foundation's EPiQS initiative through grant GBMF10104. The work at Oak Ridge National Laboratory is supported by the U.S. Department of Energy, Office of Science, National Quantum Information Science Research Centers and the Basic Energy Sciences, Materials Sciences and Engineering Division. The work at Brookhaven National Laboratory is supported by the Materials Science and Engineering Divisions, Office of Basic Energy Sciences of the U.S. Department of Energy under contract no. DESC0012704. FIB use at the Center for Functional Nanomaterials, Brookhaven National Laboratory is acknowledged.

REFERENCES

1. Qi, X.-L.; Zhang, S.-C. Topological insulators and superconductors. Rev. Mod. Phys. **2011,** 83 (4), 1057-1110.

2. Hor, Y. S.; Williams, A. J.; Checkelsky, J. G.; Roushan, P.; Seo, J.; Xu, Q.; Zandbergen, H. W.; Yazdani, A.; Ong, N. P.; Cava, R. J. Superconductivity in $Cu_xBi_2Se_3$ and its implications for pairing in the undoped topological insulator. Phys. Rev. Lett. **2010,** 104 (5), 057001.

3. Liu, Z.; Yao, X.; Shao, J.; Zuo, M.; Pi, L.; Tan, S.; Zhang, C.; Zhang, Y. Superconductivity with Topological Surface State in $Sr_xBi_2Se_3$. J. Am. Chem. Soc. **2015,** 137 (33), 10512-10515.

4. Smylie, M. P.; Claus, H.; Welp, U.; Kwok, W. K.; Qiu, Y.; Hor, Y. S.; Snezhko, A. Evidence of nodes in the order parameter of the superconducting doped topological insulator $Nb_xBi_2Se_3$ via penetration depth measurements. Phys. Rev. B **2016,** 94 (18), 180510.

5. Wang, M. X.; Liu, C.; Xu, J. P.; Yang, F.; Miao, L.; Yao, M. Y.; Gao, C. L.; Shen, C.; Ma, X.; Chen, X.; Xu, Z. A.; Liu, Y.; Zhang, S. C.; Qian, D.; Jia, J. F.; Xue, Q. K. The coexistence of superconductivity and topological order in the $Bi_2Se_3$ thin films. Science **2012,** 336 (6077), 52-55.

6. Wang, E.; Ding, H.; Fedorov, A. V.; Yao, W.; Li, Z.; Lv, Y.-F.; Zhao, K.; Zhang, L.-G.; Xu, Z.; Schneeloch, J.; Zhong, R.; Ji, S.-H.; Wang, L.; He, K.; Ma, X.; Gu, G.; Yao, H.; Xue, Q.-K.; Chen, X.; Zhou, S. Fully gapped topological surface states in $Bi_2Se_3$ films induced by a d-wave high-temperature superconductor. Nat. Phys. **2013,** 9 (10), 621-625.

7. Xu, J. P.; Wang, M. X.; Liu, Z. L.; Ge, J. F.; Yang, X.; Liu, C.; Xu, Z. A.; Guan, D.; Gao, C. L.; Qian, D.; Liu, Y.; Wang, Q. H.; Zhang, F. C.; Xue, Q. K.; Jia, J. F. Experimental detection of a Majorana mode in the core of a magnetic vortex inside a topological insulator-superconductor $Bi_2Te_3$/$NbSe_2$ heterostructure. Phys. Rev. Lett. **2015,** 114 (1), 017001.

8. Yang, H.; Li, Y. Y.; Liu, T. T.; Xue, H. Y.; Guan, D. D.; Wang, S. Y.; Zheng, H.; Liu, C. H.; Fu, L.; Jia, J. F. Superconductivity of Topological Surface States and Strong Proximity Effect in $Sn_{1-x}Pb_xTe$-Pb Heterostructures. Adv. Mater. **2019,** 31 (52), e1905582.

9. Trang, C. X.; Shimamura, N.; Nakayama, K.; Souma, S.; Sugawara, K.; Watanabe, I.; Yamauchi, K.; Oguchi, T.; Segawa, K.; Takahashi, T.; Ando, Y.; Sato, T. Conversion of a conventional superconductor into a topological superconductor by topological proximity effect. Nat. Commun. **2020,** 11 (1), 159.

10. Wan, S.; Gu, Q.; Li, H.; Yang, H.; Schneeloch, J.; Zhong, R. D.; Gu, G. D.; Wen, H.-H. Twofold symmetry of proximity-induced superconductivity in $Bi_2Te_3$/$Bi_2Sr_2CaCu_2O_{8+\delta}$ heterostructures revealed by scanning tunneling microscopy. Phys. Rev. B **2020,** 101 (22), 220503.

11. Zhang, P.; Yaji, K.; Hashimoto, T.; Ota, Y.; Kondo, T.; Okazaki, K.; Wang, Z.; Wen, J.; Gu, G. D.; Ding, H.; Shin, S. Observation of topological superconductivity on the surface of an iron-based superconductor. Science **2018,** 360 (6385), 182-186.

12. Wang, D.; Kong, L.; Fan, P.; Chen, H.; Zhu, S.; Liu, W.; Cao, L.; Sun, Y.; Du, S.; Schneeloch, J.; Zhong, R.; Gu, G.; Fu, L.; Ding, H.; Gao, H. J. Evidence for Majorana bound states in an iron-based superconductor. Science **2018,** 362 (6412), 333-335.

13. Wang, Z.; Rodriguez, J. O.; Jiao, L.; Howard, S.; Graham, M.; Gu, G. D.; Hughes, T. L.; Morr, D. K.; Madhavan, V. Evidence for dispersing 1D Majorana channels in an iron-based superconductor. Science **2020,** 367 (6473), 104-108.

14. Zhu, S.; Kong, L.; Cao, L.; Chen, H.; Papaj, M.; Du, S.; Xing, Y.; Liu, W.; Wang, D.; Shen, C.; Yang, F.; Schneeloch, J.; Zhong, R.; Gu, G.; Fu, L.; Zhang, Y. Y.; Ding, H.; Gao, H. J. Nearly quantized conductance plateau of vortex zero mode in an iron-based superconductor. Science **2020,** 367 (6474), 189-192.

15. Nie, Y. F.; Brahimi, E.; Budnick, J. I.; Hines, W. A.; Jain, M.; Wells, B. O. Suppression of superconductivity in FeSe films under tensile strain. Appl. Phys. Lett. **2009,** 94 (24), 242505.

16. Bellingeri, E.; Pallecchi, I.; Buzio, R.; Gerbi, A.; Marrè, D.; Cimberle, M. R.; Tropeano, M.; Putti, M.; Palenzona, A.; Ferdeghini, C. $T_C$=21 K in epitaxial FeSe$_{0.5}$Te$_{0.5}$ thin films with biaxial compressive strain. Appl. Phys. Lett. **2010,** 96 (10).

17. Han, Y.; Li, W. Y.; Cao, L. X.; Wang, X. Y.; Xu, B.; Zhao, B. R.; Guo, Y. Q.; Yang, J. L. Superconductivity in iron telluride thin films under tensile stress. Phys. Rev. Lett. **2010,** 104 (1), 017003.

18. Si, W.; Jie, Q.; Wu, L.; Zhou, J.; Gu, G.; Johnson, P. D.; Li, Q. Superconductivity in epitaxial thin films of Fe$_{1.08}$Te:O$_x$. Phys. Rev. B **2010,** 81 (9), 092506.

19. Li, Q.; Si, W.; Dimitrov, I. K. Films of iron chalcogenide superconductors. Rep. Prog. Phys. **2011,** 74 (12), 124510.

20. Mele, P.; Matsumoto, K.; Fujita, K.; Yoshida, Y.; Kiss, T.; Ichinose, A.; Mukaida, M. Fe–Te–Se epitaxial thin films with enhanced superconducting properties. Supercon. Sci. Technol. **2012,** 25 (8), 084021.

21. Imai, Y.; Sawada, Y.; Nabeshima, F.; Maeda, A. Suppression of phase separation and giant enhancement of superconducting transition temperature in FeSe$_{1-x}$Te$_x$ thin films. Proc. Natl. Acad. Sci. U.S.A. **2015,** 112 (7), 1937-1940.

22. Yao, X.; Brahlek, M.; Yi, H. T.; Jain, D.; Mazza, A. R.; Han, M. G.; Oh, S. Hybrid Symmetry Epitaxy of the Superconducting Fe(Te,Se) Film on a Topological Insulator. Nano Lett. **2021,** 21 (15), 6518-6524.

23. Qin, H.; Chen, X.; Guo, B.; Pan, T.; Zhang, M.; Xu, B.; Chen, J.; He, H.; Mei, J.; Chen, W.; Ye, F.; Wang, G. Moire Superlattice-Induced Superconductivity in One-Unit-Cell FeTe. Nano Lett. **2021,** 21 (3), 1327-1334.

24. Wu, J.; Liu, F.; Sasase, M.; Ienaga, K.; Obata, Y.; Yukawa, R.; Horiba, K.; Kumigashira, H.; Okuma, S.; Inoshita, T.; Hosono, H. Natural van der Waals heterostructural single crystals with both magnetic and topological properties. Sci. Adv. **2019,** 5 (11), eaax9989.

25. Hu, C.; Gordon, K. N.; Liu, P.; Liu, J.; Zhou, X.; Hao, P.; Narayan, D.; Emmanouilidou, E.; Sun, H.; Liu, Y.; Brawer, H.; Ramirez, A. P.; Ding, L.; Cao, H.; Liu, Q.; Dessau, D.; Ni, N. A van der Waals antiferromagnetic topological insulator with weak interlayer magnetic coupling. Nat. Commun. **2020,** 11 (1), 97.

26. Deng, Y.; Yu, Y.; Shi, M. Z.; Guo, Z.; Xu, Z.; Wang, J.; Chen, X. H.; Zhang, Y. Quantum anomalous Hall effect in intrinsic magnetic topological insulator $MnBi_2Te_4$. Science **2020,** 367 (6480), 895-900.

27. Koirala, N.; Brahlek, M.; Salehi, M.; Wu, L.; Dai, J.; Waugh, J.; Nummy, T.; Han, M. G.; Moon, J.; Zhu, Y.; Dessau, D.; Wu, W.; Armitage, N. P.; Oh, S. Record Surface State Mobility and Quantum Hall Effect in Topological Insulator Thin Films via Interface Engineering. Nano Lett. **2015,** 15 (12), 8245-8249.

28. Moon, J.; Koirala, N.; Salehi, M.; Zhang, W.; Wu, W.; Oh, S. Solution to the Hole-Doping Problem and Tunable Quantum Hall Effect in $Bi_2Se_3$ Thin Films. Nano Lett. **2018,** 18 (2), 820-826.

29. He, Q. L.; Liu, H.; He, M.; Lai, Y. H.; He, H.; Wang, G.; Law, K. T.; Lortz, R.; Wang, J.; Sou, I. K. Two-dimensional superconductivity at the interface of a $Bi_2Te_3$/FeTe heterostructure. Nat. Commun. **2014,** 5, 4247.

30. Liang, J.; Zhang, Y. J.; Yao, X.; Li, H.; Li, Z. X.; Wang, J.; Chen, Y.; Sou, I. K. Studies on the origin of the interfacial superconductivity of $Sb_2Te_3$/$Fe_{1+y}Te$ heterostructures. Proc. Natl. Acad. Sci. U.S.A. **2020,** 117 (1), 221-227.

31. Qin, H.; Guo, B.; Wang, L.; Zhang, M.; Xu, B.; Shi, K.; Pan, T.; Zhou, L.; Chen, J.; Qiu, Y.; Xi, B.; Sou, I. K.; Yu, D.; Chen, W. Q.; He, H.; Ye, F.; Mei, J. W.; Wang, G. Superconductivity in Single-Quintuple-Layer $Bi_2Te_3$ Grown on Epitaxial FeTe. Nano Lett. **2020,** 20 (5), 3160-3168.

32. Liang, J.; Yao, X.; Zhang, Y. J.; Chen, F.; Chen, Y.; Sou, I. K. Formation of Fe-Te Nanostructures during in Situ Fe Heavy Doping of $Bi_2Te_3$. Nanomaterials **2019,** 9 (5), 782.

33. Zhang, Z.; Cai, M.; Li, R.; Meng, F.; Zhang, Q.; Gu, L.; Ye, Z.; Xu, G.; Fu, Y.-S.; Zhang, W. Controllable synthesis and electronic structure characterization of multiple phases of iron telluride thin films. Phys. Rev. Mater. **2020,** 4 (12), 125003.

34. Liu, T. J.; Ke, X.; Qian, B.; Hu, J.; Fobes, D.; Vehstedt, E. K.; Pham, H.; Yang, J. H.; Fang, M. H.; Spinu, L.; Schiffer, P.; Liu, Y.; Mao, Z. Q. Charge-carrier localization induced by excess Fe in the superconductor $Fe_{1+y}Te_{1-x}Se_x$. Phys. Rev. B **2009,** 80 (17), 174509.

35. Li, Y.; Zaki, N.; Garlea, V. O.; Savici, A. T.; Fobes, D.; Xu, Z.; Camino, F.; Petrovic, C.; Gu, G.; Johnson, P. D.; Tranquada, J. M.; Zaliznyak, I. A. Electronic properties of the bulk and surface states of $Fe_{1+y}Te_{1-x}Se_x$. Nat. Mater. **2021,** 20 (9), 1221-1227.

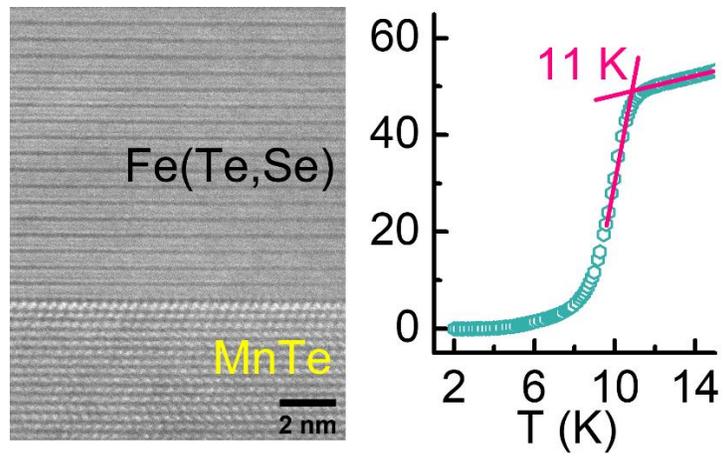

For TOC only